\documentclass[12pt]{article}
\usepackage{latexsym}
\usepackage{amsthm}
\usepackage{amsmath,amssymb,amsfonts,bm,amsbsy,bbm}
\usepackage{appendix}
\usepackage{epsfig}
\usepackage{graphicx,rotating,lscape}
\usepackage{booktabs}
\usepackage{verbatim}
\usepackage{natbib}
\usepackage{url}
\usepackage{multirow,color}
\usepackage{setspace}
\usepackage{float}
\usepackage{threeparttable}
\usepackage{algpseudocode}
\usepackage{xr-hyper}
\usepackage{hyperref}
\usepackage{xcolor}

\newtheorem{assumption}{Assumption}
\newtheorem{proposition}{Proposition}

\newtheorem{lemma}{Lemma}
\newtheorem{theorem}{Theorem}
\newtheorem{corollary}{Corollary}
\def\A{{\boldsymbol A}}
\def\X{{\boldsymbol X}}
\def\Z{{\boldsymbol Z}}
\def\W{{\boldsymbol W}}
\def\V{{\boldsymbol V}}
\def\U{{\boldsymbol U}}
\def\I{{\boldsymbol I}}
\def\M{{\boldsymbol M}}
\def\C{{\boldsymbol C}}
\def\T{{\boldsymbol T}}

\def\bSigma{{\boldsymbol \Sigma}}
\def\bOmega{{\boldsymbol \Omega}}
\def\a {{\boldsymbol a}}
\def\y {{\boldsymbol y}}
\def\balpha{{\boldsymbol \alpha}}
\def\beps {{\boldsymbol \epsilon}}

\def\bmu {{\boldsymbol \mu}}
\def\beeta {{\boldsymbol \eta}}
\def\bxi {{\boldsymbol \xi}}

\def\bgamma {{\boldsymbol \gamma}}
\def\bDelta {{\boldsymbol \Delta}}

\newcommand{\R}{\mathbb{R}}
\newcommand{\simiid}{\stackrel{\textnormal{i.i.d.}}{\sim}}

\def\s{\sigma}
\def\t{\top}
\def\sumjp{\sum^p_{j=1}}
\def\sumj{\sum^{p-1}_{j=1}}
\def\eps{\epsilon}
\def\l{\left}
\def\r{\right}

\definecolor{maroon(html/css)}{rgb}{0.5, 0.0, 0.0}

\newcommand{\blind}{1}

\makeatletter
\newcommand*{\addFileDependency}[1]{
\typeout{(#1)}
\@addtofilelist{#1}
\IfFileExists{#1}{}{\typeout{No file #1.}}
}\makeatother

\begin{document}

\def\spacingset#1{\renewcommand{\baselinestretch}%
{#1}\small\normalsize} \spacingset{1}


\if1\blind
{
  \title{\bf Debiased high-dimensional regression calibration for errors-in-variables log-contrast models}
  \author{Huali Zhao\\
    Department of Mathematical Sciences, Tsinghua University\\
    and \\
    Tianying Wang\thanks{Tianying.Wang@colostate.edu}\hspace{.2cm}\\
    Department of Statistics, Colorado State University}
  \maketitle
} \fi

\if0\blind
{
  \bigskip
  \bigskip
  \bigskip
  \begin{center}
    {\LARGE\bf Debiased high-dimensional regression calibration for errors-in-variables log-contrast models}
\end{center}
  \medskip
} \fi

\bigskip

\begin{abstract}
Motivated by the challenges in analyzing gut microbiome and metagenomic data, this work aims to tackle the issue of measurement errors in high-dimensional regression models that involve compositional covariates. This paper marks a pioneering effort in conducting statistical inference on high-dimensional compositional data affected by mismeasured or contaminated data. We introduce a calibration approach tailored for the linear log-contrast model. Under relatively lenient conditions regarding the sparsity level of the parameter, we have established the asymptotic normality of the estimator for inference. Numerical experiments and an application in microbiome study have demonstrated the efficacy of our high-dimensional calibration strategy in minimizing bias and achieving the expected coverage rates for confidence intervals. Moreover, the potential application of our proposed methodology extends well beyond compositional data, suggesting its adaptability for a wide range of research contexts.
\end{abstract}

\noindent%
\vfill

\newpage
\spacingset{1.9} 

\section{Introduction}\label{sec:introduction} 
Compositional data, which include proportions or percentages that sum to a unit, are pivotal in diverse cross-disciplinary studies. Examples include the analysis of food nutrient compositions in nutrition, the examination of asset portfolio compositions in finance, and the investigation of rock geochemical compositions in geochemistry. The advent of large and high-dimensional compositional datasets poses a challenge to classical high-dimensional methods, primarily due to the oversight of the unit sum constraint. Consequently, there is a concerted effort among researchers to devise statistical methodologies tailored for high-dimensional compositional data analysis, with a notable emphasis on genomic studies \citep{xia2013logistic, lin2014variable, wang2017structured}. This paper endeavors to model high-dimensional compositional data while concurrently addressing the pervasive issue of measurement errors, a concern that spans multiple disciplines \citep{dembicki2009three, escoffier2015quantifying}. Prior works have identified such circumstances as the issue of errors-in-variables, demonstrating that applying standard regression techniques directly to mismeasured covariates can lead to imprecise inferential outcomes \citep{hausman2001mismeasured, shi2022high}.

Hardly any methods have been developed to address measurement error problems with high-dimensional compositional data. The fact that the composition components must sum to a unit renders common analysis methods for measurement error inappropriate or inapplicable. In the field of low-dimensional compositional data analysis, \citet{aitchison1984measurement} firstly established the measurement error model for compositional data and assessed the effect of measurement errors on hypothesis testing of complete subcompositional independence about the distribution of a composition. However, their work did not address statistical inferences on regression problems involving mismeasured compositional covariates. In high-dimensional regression analysis (i.e., large number of variables, small sample size, $p\gg n$), the analysis of measurement errors is also limited. Both \citet{loh2012high} and \citet[CoCoLasso]{datta2017cocolasso} proposed methods based on the Lasso \citep{tibshirani1996regression} from the prediction perspective for general high-dimensional regression with measurement errors and missing data. In microbial compositional data analysis, \citet{shi2022high} highlighted the measurement errors introduced by imputing excessive zero counts with non-zero values and provided a strategy to minimize the influence of imputing zero counts in estimating coefficients for modeling compositional data transferred from imputed count data. However, they overlook the measurement errors in non-zero counts due to one-time observations \citep{jiang2023flexible}. Thus, there is a lack of methods for statistical inference in high-dimensional regression with mismeasured compositional covariates.

Log-contrast models were introduced by \citet{aitchison1984log} for modeling log contrasts of the proportions from experiments with mixtures and have been proven useful for a wide variety of regression problems \citep{wang2017structured,li2022s,han2023robust}. For example, \citet{wang2017structured} employed a log-contrast model to explore the relationship between the body mass index (BMI) of an individual and the composition of the gut microbiome a person carries. Suppose that we observe an $n$-vector $\y=(y_1,\ldots,y_n)^\t$ representing the responses of $n$ individuals (e.g., BMI) and an $n\times p$ matrix $\X=(x_{ij})$ of $p$ variables of $n$ individuals (e.g., count data of microbial taxa; food nutrient components), which is normalized into an $n\times p$ compositional data matrix $\Z=(z_{ij})$ with $z_{ij} = x_{ij}/(\sumjp x_{ij})$. It is more reasonable to model the relationship between $y_i$ and $\Z_i=(z_{i1},\ldots,z_{1p})^\t$ as $\Z_i$ is scale-free. However, the transformation from $\X_i$ to $\Z_i$ introduces a unit-sum constraint, causing non-identifiability in classical regression models. To address this challenge, \citet{aitchison1984log} proposed applying the log-ratio transformation \citep{aitchison1982statistical} to compositional covariates $\Z_i$ and constructing the linear log-contrast model
\begin{equation}\label{eq:logcontrastZ}
    y_i = \sumj \alpha_j\log\left(\frac{z_{ij}}{z_{ip}}\right) + \eps_i \quad (i=1,\ldots, n),
\end{equation}
where $\balpha=(\alpha_1,\ldots,\alpha_{p-1})^\t \in \R^{p-1}$ is the regression coefficient vector, and $\beps=(\eps_1,\ldots,\eps_n)^\t$ is an $n$-vector of independent noise distributed as $N(0,\s^2_{\eps})$. Model \eqref{eq:logcontrastZ} does not contain an intercept term, as it is eliminated by centering both responses and predictor variables \citep{lin2014variable,shi2016regression}. By setting $\alpha_p=-\sumj\alpha_j$, model~\eqref{eq:logcontrastZ} is equivalent to $y_i = \sumjp \alpha_j\log(z_{ij}) + \eps_i$ with a constraint $\sumjp \alpha_j=0$. Both formulas are common in microbiome compositional data analysis \citep{shi2016regression,wang2017structured}. However, it has been pointed out that detecting microbial associations with complex diseases, such as cardiovascular disease, remains a daunting challenge due to the complexity of microbiome data and a lack of appropriate statistical methods \citep{xu2021zero}. The zero-sum constraint of the coefficients may cause complications in establishing measurement error models and solving the model subsequently. 

Much of the literature considers variable selection and estimation in the high-dimensional regression with compositional covariates based on model \eqref{eq:logcontrastZ} but without measurement errors \citep{lin2014variable,shi2016regression,wang2017structured}. However, in 
many applications of nutrition, econometrics, geochemistry, and microbiology, we also frequently see covariates corrupted with noise \citep{aitchison1984statistical,shi2022high,jiang2023flexible}. When we do not observe $\X$, but its contaminated version $\W=(w_{ij})\in\R^{n\times p}$, the $n\times p$ compositional data matrix with measurement errors is $\V=(v_{ij})$ with $v_{ij} =w_{ij}/(\sumjp w_{ij})$. A naive method that replaces true compositions $\Z$ by observed compositions $\V$ can be written as follows
\begin{equation}\label{eq:logcontrastV}
    y_i = \sumj \widetilde{\alpha}_j\log\left(\frac{v_{ij}}{v_{ip}}\right) + \widetilde\eps_i \quad (i=1,\ldots, n),
\end{equation}
where $\widetilde{\balpha}=(\widetilde{\alpha}_1,\ldots,\widetilde{\alpha}_{p-1})^\t\in \R^{p-1}$ is the regression coefficient vector, and $\widetilde\beps=(\widetilde\eps_1,\ldots,\widetilde\eps_n)^\t$ is an $n$-vector of independent noise distributed as $N(0,\s^2_{\widetilde{\epsilon}})$. Building on the framework of \citet[CoCoLasso]{datta2017cocolasso} with data generated by model \eqref{eq:logcontrastV} and our subsequent measurement error model \eqref{eq:mem_composition}, we can estimate coefficients and conduct variable selection effectively. However, what remains unaddressed is how to perform inference in the presence of measurement errors, which is crucial for application, especially in microbiome studies. For example, \citet{aitchison1984measurement} highlighted the significance of understanding how inaccuracies in measuring compositional data affect statistical inferences. Moreover, recent findings by \citet{gihawi2023major} have also shown that errors in the microbial reads and the transformation of raw data invalidate cancer microbiome findings. This suggests the growing necessity to develop robust statistical inference methods that can account for the complexities of high-dimensional data with measurement errors. 

In this paper, we focus on the challenges posed by measurement errors within the framework of model~\eqref{eq:logcontrastZ}, which is widely used in microbiome studies \citep{lin2014variable,wang2017structured}. Our primary goal is to conduct statistical inference for the relative coefficient $\alpha_j$, which expresses the effect of the $j$th component relative to the $p$th component on the response. We comprehensively discuss the challenges of measurement error problems in high-dimensional compositional data and provide satisfactory solutions for its statistical inference. We propose a pioneering approach for statistical inference with contaminated 
high-dimensional compositional data, named the high-dimensional calibration method. The asymptotic normality of the proposed estimator is established under mild conditions on the sparsity level of the parameter. 

\section{Methodology}\label{sec:methodology}
\subsection{Notations and measurement error models}
Let $\boldsymbol{1}_p$ be a $p$-vector with all elements as $1$ and $\I_p$ be a $p$-order identity matrix for any positive integer $p$. For any vector $\a=(a_1,\ldots,a_p)^\t$ and positive integer $q$, we write $\|\a\|_q$ for the standard $l_q$ norm, i.e., $\|\a\|_q=(\sumjp |a_j|^q)^{1/q}$, and write $\|\a\|_{\infty}$ for the maximum norm, i.e., $\|\a\|_{\infty}=\max_{1\le j\le p}|a_j|$. For ease of representation, we write $a_j$ as $\a[j]$ and write $\log(\a)=(\log(a_1),\ldots,\log(a_p))^\t$. For a matrix $\A=(a_{jj'})\in\R^{p\times p}$ and any positive integer $q$, $\|\A\|_q$ denotes its $l_q$ operator norm; in particular, $\|\A\|_{\infty}=\max_{1\le j\le p}\sum^p_{j'=1}|a_{jj'}|$. This is to be contrasted with the maximum absolute of any entry of $\A$, denoted by $|\A|_{\infty}=\max_{1\le j,j'\le p}|a_{jj'}|$. The maximum and the minimum singular values of a matrix $\A$ are denoted by $\sigma_{\max}(\A)$ and $\sigma_{\min}(\A)$, respectively. For simplicity, we also write $a_{jj'}$ as $\A[j,j']$ or $\A_{j,j'}$ and write $\log(\A)=(\log(a_{jj'}))\in\R^{p\times p}$. Define $s_0$ as the number of nonzero entries of $\balpha$ that is defined in Eq~\eqref{eq:logcontrastZ}. Denote $C_{\min}$ and $C_{\max}$ as two positive constants. For two functions $f(n),g(n)$, we write $f(n)=o(g(n))$ if $\lim_{n\to\infty}f(n)/g(n)=0$ and write $f(n)=O(g(n))$ if $f(n)\le Cg(n)$ for some positive constant $C$. We use $o_P(\cdot),O_P(\cdot)$ to indicate asymptotic behavior in probability as the sample size $n$ approaches infinity. The symbol $\simiid$ denotes sampling independent and identically distributed random variables. Without the loss of generality, we assume that the $n$ observations of any variable are independent and identically distributed throughout the paper. A summary of the common notations used in this paper is provided in Appendix Section~1 (Table~A1). 

Multiplicative errors are particularly suitable to model the nonnegative components \citep{aitchison1984measurement,wei2009quantile}. In the context of the log-contrast models \eqref{eq:logcontrastZ} and \eqref{eq:logcontrastV}, multiplicative errors between $\W$ and $\X$, see Eq~\eqref{eq:mem}, can be converted to additive errors between the log-contrasts of compositional data $\widetilde{\V}_{i,-p}$ and $\widetilde{\Z}_{i,-p}$, see Eq~\eqref{eq:mem_composition}, which are commonly assumed in the literature \citep{wei2009quantile,achic2018categorizing}.
Specifically, we consider the multiplicative lognormal error model
\begin{equation}\label{eq:mem}
    w_{ij} = x_{ij}u_{ij},\quad \log(u_{ij}) \simiid N(-0.5\sigma^2_u,\sigma^2_u),
\end{equation}
where the measurement error $\U=(u_{ij})\in\R^{n\times p}$ is independent of $\X$. The mean of $\log(u_{ij})$ is chosen to ensure that $\W$ is an unbiased measure of $\X$, i.e., $E(\W\mid \X)=\X$, by setting $E(u_{ij})=1$ \citep[Chapter 4.5.2]{carroll2006measurement}.
For simplicity, we assume that the variance of measurement errors from sampling is the same across different components. One can assume $\log(u_{ij}) \sim N(-0.5\sigma^2_{uj},\sigma^2_{uj})$ for the $j$th component when different components have different measurement errors if there are sufficient replicates for estimating $\sigma^2_{uj}$. 

    One point worth noting is that the identifiability of the parameter vector $\balpha\in\R^{p-1}$ does not depend on the data distribution in outcome models \eqref{eq:logcontrastZ}-\eqref{eq:logcontrastV}, it only depends on the identifiability of measurement errors. Under our multiplicative lognormal error model \eqref{eq:mem}, the key is to identify the variance of $\log(u_{ij})$, $\s^2_u$, which might be unknown in practice and can be estimated based on repeated observations of $\W$ or external data (see Eq~\eqref{eq:sigma_u}, Section~\ref{sec:implement}) in the literature \citep{aitchison1984log,achic2018categorizing}. In a more general case, if there are no assumptions of the distributional form of measurement errors, one needs to assume the density function of measurement errors is known or can be estimated consistently in practice.

To introduce the main model, we denote $\widetilde{\Z}_{i,-p}\in \R^{p-1}$ and $\widetilde{\V}_{i,-p}\in \R^{p-1}$ as below
\begin{align*}
    \widetilde{\Z}_{i,-p}=\left(\log\l(\frac{z_{i1}}{z_{ip}}\r),\ldots,\log\l(\frac{z_{i,p-1}}{z_{ip}}\r)\right)^\t ,\quad
    \widetilde{\V}_{i,-p}=\left(\log\l(\frac{v_{i1}}{v_{ip}}\r),\ldots,\log\l(\frac{v_{i,p-1}}{v_{ip}}\r)\right)^\t.   
\end{align*}
Then, model \eqref{eq:logcontrastZ} can be written as $y_i=\balpha^\t\widetilde{\Z}_{i,-p}+\epsilon_i$, and model \eqref{eq:logcontrastV} can be written as $y_i=\widetilde\balpha^\t\widetilde{\V}_{i,-p}+\widetilde\epsilon_i$. 
Based on model \eqref{eq:mem}, and the facts of $\log(z_{ij}/z_{ip})=\log(x_{ij}/x_{ip})$ and $\log(v_{ij}/v_{ip})=\log(w_{ij}/w_{ip})$, we have 
\begin{eqnarray}\label{eq:mem_composition}
    \widetilde{\V}_{i,-p} = \widetilde{\Z}_{i,-p} + \widetilde{\U}_{i,-p},
\end{eqnarray}
where $\widetilde{\U}_{i,-p}=(\log(u_{i1}/u_{ip}),\ldots,\log(u_{i,p-1}/u_{ip}))^\t\in\R^{p-1}$ with $\log(u_{ij}/u_{ip})\sim N(0,2\s^2_u)$ for $j=1,\ldots,p-1$. 
One can also consider a measurement error model based on compositional data directly, which implies assuming Eq~\eqref{eq:mem_composition} instead of Eq~\eqref{eq:mem} in the context of the linear log-contrast model \eqref{eq:logcontrastV}, and our proposed method subsequently is still valid. We establish model \eqref{eq:mem} directly rather than model \eqref{eq:mem_composition} considering that measurement errors actually occur at unnormalized data $\X$ rather than compositional data $\Z$ \citep{jiang2023flexible}.

\subsection{The high-dimensional calibration method}\label{subsec:HigCal}
In the analysis of compositional data, it is often a standard practice to perform a log transformation, particularly in microbiome studies, due to the initially large scale of the original compositions \citep{aitchison1984log,shi2016regression,lin2014variable}. Thus, it is more straightforward to make assumptions on the log-transformed data: we assume the log-contrast vector of compositional data $\Z_i$, namely $\widetilde{\Z}_{i,-p}$, follows a lognormal distribution. 

\begin{assumption}\label{assum-Z}
Assume $\widetilde{\Z}_{i,-p}$ follows a multivariate normal distribution with the mean vector $\bmu_{\widetilde z}\in\R^{p-1}$ and the covariance matrix $\bSigma_{\widetilde z}\in \R^{(p-1)\times (p-1)}$.
\end{assumption}
This assumption is weaker than the commonly employed assumption in the analysis of compositional data that $\Z_i$ follows a lognormal distribution \citep{aitchison1984measurement}.

\begin{proposition}\label{prop:logzv} 
Under Assumption \ref{assum-Z}, we have
$
    \widetilde{\V}_{i,-p} \simiid N_{p-1}(\bmu_{\widetilde z},\bSigma_{\widetilde z}+2\s^2_u\I_{p-1}),
$
and the conditional distribution of $\widetilde{\Z}_{i,-p}$ given $\widetilde{\V}_{i,-p}$ is Gaussian with the mean vector
\begin{eqnarray}\label{eq:rcnormal}
     \bmu(\V_i) = \bmu_{\widetilde z} + \bSigma_{\widetilde z} (\bSigma_{\widetilde z}+2\s^2_u\I_{p-1})^{-1}(\widetilde{\V}_{i,-p} - \bmu_{\widetilde z})\in\R^{p-1},
\end{eqnarray}
and the covariance matrix $\bSigma_{\widetilde z} - \bSigma_{\widetilde z} (\bSigma_{\widetilde z}+2\s^2_u\I_{p-1})^{-1}\bSigma_{\widetilde z}$. Here, $\bmu(\V_i)$ depends on $\widetilde{\V}_{i,-p}$, which is intrinsically a function of $\V_i=(v_{i1},\ldots,v_{ip})^\top$ based on its definition.
Further, the covariance matrix of $\bmu(\V_i)$ is
\begin{eqnarray}\label{eq:Sigma}
    \bSigma = \bSigma_{\widetilde z}(\bSigma_{\widetilde z}+2\s^2_u\I_{p-1})^{-1}\bSigma_{\widetilde z}.
\end{eqnarray}  
\end{proposition}
In classical settings with $n\gg p$, one common approach for measurement error correction is regression calibration \citep{carroll1990approximate,gleser1990improvements}, which replaces 
$\widetilde{\V}_{i,-p}$ in model \eqref{eq:logcontrastV} with the conditional mean of $\widetilde{\Z}_{i,-p}$ given $\widetilde{\V}_{i,-p}$, that is, $\bmu(\V_i)$ defined in Eq~\eqref{eq:rcnormal}. Then, the calibration model with observed compositional covariates is
\begin{equation}\label{eq:rcmodel}
    y_i = \balpha^\t\bmu(\V_i) + \eta_i\quad (i=1,\ldots,n),
\end{equation}
where $\beeta=(\eta_1,\ldots,\eta_n)^\t$ is an $n$-vector of independent Gaussian noise with mean zero and variance $\s^2$. Unfortunately, classical regression calibration methods, relying on ordinary least squares, fail in high-dimensional cases (i.e., $p\gg n$). As the first attempt to construct inference for high-dimensional compositional data with measurement errors, we next consider a debiased-Lasso-type estimator based on model \eqref{eq:rcmodel} and derive its asymptotic distribution. 

Given $n$ observations $(y_i,\W_i)^n_{i=1}$ with $y_i\in\R,\W_i\in\R^p$, we first assume all nuisance parameters $\s^2_u$, $\bmu_{\widetilde z}$, and $\bSigma_{\widetilde z}$ are known, which indicates that conditional mean $\bmu(\V_i)$ is deterministic based on Eq~\eqref{eq:rcnormal} with $\V_i$ determined by $\W_i$. We provide detailed implementation and estimation of nuisance parameters $\s^2_u$, $\bmu_{\widetilde z}$, and $\bSigma_{\widetilde z}$ in the context of microbiome studies in Section~\ref{sec:implement}. Define the matrix $\M(\V)\in\R^{n\times(p-1)}$ with its $i$th row being $\bmu(\V_i)$. Without loss of generality, we assume $\y=(y_1,\ldots,y_n)^\t$ is scaled to have empirical mean zero and $\M(\V)$ is scaled to have row-mean zero throughout the paper. Then, the rows of $\M(\V)$ are mutually independent with normal distribution $N_{p-1}(\boldsymbol{0},\bSigma)$, where $\bSigma$ is defined in Eq~\eqref{eq:Sigma}. Then, we rewrite the calibration model \eqref{eq:rcmodel} as the following matrix form
\begin{equation}\label{eq:rcmodel_matrix}
    \y = \M(\V)\balpha + \beeta, \quad \beeta \sim N_n(\boldsymbol{0},\s^2 \I_n).
\end{equation}
To mitigate the difficulties associated with high dimensionality when $p\gg n$, we estimate $\balpha$ through the Lasso by solving the following convex optimization problem
\[
    \widetilde\balpha(\y,\M(\V);\lambda) = \arg\min_{\balpha\in \R^{p-1}} \left\{\frac{1}{2n}\|\y-\M(\V)\balpha\|^2_2 + \lambda\|\balpha\|_1\right\},
\]
where $\lambda$ is a regularization parameter and is often selected by cross-validation. With a slight abuse of notation, we use $\widetilde\balpha$ instead of $\widetilde\balpha(\y,\M(\V);\lambda)$ if no ambiguity is raised in the context. Then, we propose a debiased-Lasso type estimator, namely the high-dimensional calibration estimator,
\begin{equation}\label{eq:Proposed}
    \widehat\balpha = \widetilde\balpha + \frac{1}{n}\bSigma^{-1}\M(\V)^\t(\y-\M(\V)\widetilde\balpha),
\end{equation}
and establish its theoretical properties in Section~\ref{sec:theory}. The proposed estimator is asymptotically unbiased and normal under mild conditions on the sparsity level of the parameter, which is important for statistical inference urgently needed in many applications. For example, $\alpha_j$ reflects the influence of the relative abundances of gut microbiomes on BMI in microbiome studies. The valid inference for $\alpha_j$ could assist researchers in comprehending the specific connections between microbiomes and various health problems associated with obesity, such as type 2 diabetes, high blood pressure, and heart disease. 

\subsection{Implementation and estimation in microbiome studies}\label{sec:implement}
The analysis of microbiome compositional data plays a major role in exploring the relationship between phenotypes and the composition of the microbiome within the human body. However, measurement errors are common in microbiome data, arising from limitations like single-time-point observations and batch effects~\citep{jiang2023flexible,dai2019batch}. In this section, we will outline the application of the proposed method in the context of microbiome research, focusing on the estimation of nuisance parameters $\s^2_u,\bmu_{\widetilde z}$, and $\bSigma_{\widetilde z}$. 

In the analysis of microbiome data, it is common to assume $\log(\X_i)$ follows a multivariate normal distribution with the mean $p$-vector $\bmu_x$ and the covariance matrix $\bSigma_x\in \R^{p\times p}$, which is also common and convenient for modeling positive measurements in various fields~\citep{wang2017structured,li2022s}. Then, Assumption \ref{assum-Z} naturally holds. Based on Eq~\eqref{eq:mem}, we have
$
    \log(\W_i) \simiid N_p(\bmu_w=\bmu_x-0.5\s^2_u\boldsymbol{1}_p,\bSigma_w=\bSigma_x+\s^2_u\I_p).
$
Based on the definition of $\widetilde{\Z}_{i,-p}$, we have
$\widetilde{\Z}_{i,-p} \simiid N_{p-1}(\bmu_{\widetilde z},\bSigma_{\widetilde z})$ with $\bmu_{\widetilde z}[j] = {\bmu_x}[j]-{\bmu_x}[p]$ and
$
    \bSigma_{\widetilde z}[j,j'] = {\bSigma_x}[j,j'] - {\bSigma_x}[j,p] - {\bSigma_x}[p,j'] + {\bSigma_x}[p,p],
$
for each $j,j'\in\{1,\ldots,p-1\}$. That is, nuisance parameters $\bmu_{\widetilde z},\bSigma_{\widetilde z}$ are determined by $\bmu_x,\bSigma_x$, respectively. 

The estimation of nuisance parameters $\s^2_u, \bmu_{\widetilde z},\bSigma_{\widetilde z}$ is equivalent to the estimation of $\s^2_u, \bmu_x,\bSigma_x$. In practice, $\s^2_u$ is estimated by constructing the estimating equation based on the multiplicative error model~\eqref{eq:mem} and repeated observations of $\W$, which are commonly required in the literature \citep{aitchison1984log,achic2018categorizing}. Suppose we have $R$ replicates $\W^{(r)}=(w^{(r)}_{ij})\in\R^{n\times p},r=1,\ldots,R$ of $\W=(w_{ij})$. Define $\widehat\s^2_{u,ij}$ as the sample variance of the $\log(w^{(r)}_{ij})$ to identify $\s^2_u$, then the estimating equation for $\s^2_u$ is
\begin{eqnarray}\label{eq:sigma_u}
    (np)^{-1}\sum^n_{i=1}\sum^p_{j=1}(\widehat\s^2_{u,ij}/R-\s^2_u/R) = 0.        
\end{eqnarray}
Then, a consistent estimator of $\bmu_x$ based on Eq~\eqref{eq:mem} is 
\begin{eqnarray}\label{eq:mu_x}
    \widehat\bmu_x = \widehat\bmu_w + 0.5\widehat\s^2_u\boldsymbol{1}_p,
\end{eqnarray}
where $\widehat\bmu_w\in\R^p$ is the sample mean of $\log(\W_i)$, that is, $\widehat\bmu_w[j] = (\sum^n_{i=1}\log(w_{ij}))/n$. 
 
We now provide two methods for estimating $\bSigma_x$ without additional structure to accommodate different application scenarios. First, we could construct the node-wise estimator of covariance matrix $\bSigma_x$, following \citet{van2014asymptotically,javanmard2018debiasing}. In this paper, we describe this construction based on observed $\W$ and refer readers to \citet{javanmard2018debiasing} for more methodological details. For $j\in\{1,\ldots,p\}$, we define the vector $\widehat\bgamma_j=(\widehat\gamma_{jj'})\in\R^{p-1},j'\in\{1,\ldots,p\},j'\ne j$ by performing sparse regression of the $j$th column of $\log(\W)$ against all the other columns. Specifically,
\[
    \widehat\bgamma_j = \arg\min_{\bgamma\in\R^{p-1}}\left\{\frac{1}{2n}\|\log(\W_j)-\log(\W_{-j})\bgamma\|^2_2 + \widetilde\lambda\|\bgamma\|_1\right\},
\]
where $\log(\W_{-j})$ is the submatrix obtained by removing the $j$th column of $\log(\W)$ and $\widetilde\lambda$ is a regularization parameter. Let $\widehat\C$ be a $p\times p$ matrix, where all diagonal elements are one, and off-diagonal elements are $-\widehat\gamma_{jj'}$ for all $j\ne j'$, and let 
\[
    \widehat\T^2 = diag(\widehat\tau^2_1,\ldots,\widehat\tau^2_p),\quad \widehat\tau^2_j = \frac{1}{n}\l(\log(\W_j)-\log(\W_{-j})\widehat\bgamma_j\r)^\t\log(\W_j).
\]
Then, the node-wise estimator of covariance matrix $\bSigma_x$ is defined as
\begin{equation}\label{eq:nodewise}
    \widehat\bSigma_{x,nw} = \widehat\C^{-1}\widehat\T^2 - \widehat\s^2_u\I_p.
\end{equation}

One can also estimate $\bSigma_x$ based on the shrinkage approach proposed by \citet{schafer2005shrinkage,opgen2007accurate}. We first obtain the shrinkage estimator $\widehat\bSigma_{w,sh}$ of $\bSigma_w$ based on \citet{corpcor2021}. 
Then, the shrinkage estimator of covariance matrix $\bSigma_x$ is defined as
\begin{equation}\label{eq:shrinkage}
    \widehat\bSigma_{x,sh} = \widehat\bSigma_{w,sh} - \widehat\s^2_u\I_p.
\end{equation}
With estimated nuisance parameters $\widehat\s^2_u,\widehat\bmu_x,\widehat\bSigma_x$ ($\widehat\bSigma_{x,nw}$ or $\widehat\bSigma_{x,sh}$), based on Eq~\eqref{eq:rcnormal}, the estimation of $\mu(\V_i)$ is
$
    \widehat\bmu_{\widetilde z} + \widehat\bSigma_{\widetilde z} ( \widehat\bSigma_{\widetilde z}+2\widehat\s^2_u\I_{p-1})^{-1}(\widetilde{\V}_{i,-p} - \widehat\bmu_{\widetilde z}),
$
where $\widehat\bmu_{\widetilde z}[j] = {\widehat\bmu_x}[j]-{\widehat\bmu_x}[p]$ and 
$
    \widehat\bSigma_{\widetilde z}[j,j'] = {\widehat\bSigma_x}[j,j'] - {\widehat\bSigma_x}[j,p] - {\widehat\bSigma_x}[p,j'] + {\widehat\bSigma_x}[p,p],
$
for each $j,j'\in\{1,\ldots,p-1\}$. Based on Eq~\eqref{eq:Sigma}, the estimation of $\bSigma$ is 
$
    \widehat\bSigma_{\widetilde z}(\widehat\bSigma_{\widetilde z}+2\widehat\s^2_u\I_{p-1})^{-1}\widehat\bSigma_{\widetilde z}.
$
Then, we can implement our proposed method in Section~\ref{subsec:HigCal} and obtain the point estimator and confidence intervals of $\balpha$. 
    From a practical perspective, $\widehat\bSigma_{x,nw}$ is more suitable for large samples and strong correlation cases (e.g., $n=200,\rho=0.5$ in our simulation) since it can capture the correlation based on the Gaussian graphical model and the Lasso, requiring a larger sample size for accurate results. The shrinkage estimator $\widehat\bSigma_{x,sh}$ defined in Eq~\eqref{eq:shrinkage} is shrunk towards the diagonal matrix, leading to a bias of non-diagonal elements. This also suggests that $\widehat\bSigma_{x,sh}$ will perform better than $\widehat\bSigma_{x,nw}$ in small sample or weak correlation cases (e.g., $n=100,\rho=0.2$ in our simulation), which is consistent with our simulation results in Section~\ref{sec:simulation}. In practice, the choice of an appropriate estimator can be guided by the specific structure of the data at hand, with various strategies available as outlined in the literature \citep{ledoit2004well,bickel2008regularized,cai2012optimal}. While a detailed exploration of these considerations is beyond the scope of this paper, it is important to recognize the significance of selecting an estimator that aligns with the data's characteristics.


\subsection{Connection with other methods}
In this section, we discuss how the proposed method relates to and differentiates from existing approaches for high-dimensional data. A summary of the applicable situations and the features of these methods are presented in Table~\ref{tab:methods}.

The Lasso method can be applied for the log-contrast model $y_i=\widetilde\balpha^\t\widetilde{\V}_{i,-p}+\epsilon_i,\ i=1,\ldots,n$, to conduct variable selection and to improve prediction accuracy and interpretability of the model at the expense of a little bias. Several debiased Lasso estimators have been proposed to address the inference problems for the high-dimensional regression
\citep{van2014asymptotically,javanmard2018debiasing,li2020debiasing}. However, these methods do not account for measurement errors, and thus, applying the Lasso or debiased Lasso to estimate $\balpha$ in model \eqref{eq:logcontrastZ} can result in severe bias with invalid confidence intervals (see Tables~\ref{tab:n200rho0.2}-\ref{tab:n100rho0.2} in Section~\ref{sec:simulation}). 

The CoCoLasso, improved the method of \citet{loh2012high}, could handle the additive measurement error $\widetilde{\V}_{i,-p} = \widetilde{\Z}_{i,-p} + \widetilde{\U}_{i,-p}$ for log-contrasts of compositional data in high-dimensional sparse linear regression. Specifically, the method estimates the sample covariance of 
$\widetilde{\Z}_{i,-p}$ using $\widehat\bSigma_{\widetilde z} = \widehat\bSigma_{\widetilde v} - \bSigma_{\widetilde u}$, where $\widehat\bSigma_{\widetilde v}$ is sample covariance of $\widetilde{\V}_{i,-p}$, and $\bSigma_{\widetilde u}$, assumed to be known, represents the covariance of 
$\widetilde{\U}_{i,-p}$. This estimate, $\widehat\bSigma_{\widetilde z}$, is then replaced with its nearest positive semi-definite matrix of $\widehat\bSigma_{\widetilde z}$ in the optimization problem:
$
    \arg\min_{\balpha\in\R^{p-1}}\frac{1}{2}\balpha^\t\widehat\bSigma_{\widetilde z}\,\balpha - \frac{1}{n}\sum^n_{i=1}y_i(\widetilde{\V}_{i,-p}^\t\balpha) + \lambda^* \|\balpha\|_1
$
to estimate $\balpha$, where $\lambda^*$ is a regularization parameter. While CoCoLasso performs well in prediction and variable selection, it lacks the capability for statistical inference on high-dimensional regularized estimator of $\balpha$. 

Unlike CoCoLasso, our method replaces the contaminated covariates $\widetilde\V_{i,-p}$ with calibrated covariates $\bmu(\V_i)$, which serves as an unbiased approximation of true log-contrasts $\widetilde\Z_{i,-p}$, ensuring that $E\{\bmu(\V_i)\} = E(\widetilde\Z_{i,-p})$. We propose a debiased-Lasso type estimator Eq~\eqref{eq:Proposed} to facilitate statistical inference for $\balpha$. Unlike the common debiased Lasso estimator, our covariates $\bmu(\V_i)$, which contain inherent randomness, cannot be directly observed like $\widetilde\V_{i,-p}$ because they depend on nuisance parameters $\s^2_u,\bmu_{\widetilde z},\bSigma_{\widetilde z}$ due to the presence of measurement errors. In addition, the proposed method establishes a multiplicative lognormal error model, which is general and flexible enough to incorporate a wide range of measurement error problems. Moreover, the proposed method can also handle general high-dimensional regression with additive measurement errors, under mild conditions, not limited to compositional data; see Appendix Section~2 
for a detailed description.

\begin{table}
\begin{center}
\caption{Summary of four different methods. DeLasso, debiased Lasso method; Proposed, proposed high-dimensional calibration method.}\label{tab:methods}
\begin{tabular}{lcccc}
\hline
Method & $p\gg n$ & Compositional data & Measurement errors & Inference \\
\hline
Lasso           & \checkmark & $\times$     & $\times$     & $\times$  \\  
DeLasso         & \checkmark & $\times$     & $\times$     & \checkmark \\
CoCoLasso       & \checkmark & $\times$     & \checkmark   & $\times$   \\
Proposed        & \checkmark & \checkmark   & \checkmark   & \checkmark \\
\hline
\end{tabular}
\end{center}
\end{table}

\section{Theoretical Properties}\label{sec:theory}
We now investigate the theoretical performance of the proposed high-dimensional calibration estimator for the log-error-in-variable regression with compositional covariates. First, we show that under the nearly optimal condition of sparsity level $s_0=o(n/(\log(p-1))^2)$, the proposed estimator $\widehat\balpha$ in Eq~\eqref{eq:Proposed} is asymptotically Gaussian.
\begin{proposition}\label{prop:Proposed}
    Define the high-dimensional calibration estimator $\widehat\balpha$ via Eq~\eqref{eq:Proposed} with $\lambda = 8\s\sqrt{(\log(p-1))/n}$. If $n,p\to\infty$ with $s_0=o(n/(\log(p-1))^2)$, then, setting $\bxi = \bSigma^{-1/2}\M(\V)^\top\beeta/\sqrt{n}$, we have
    \begin{equation}\label{eq:asyProposed}
        \sqrt{n}(\widehat\balpha-\balpha) = \bxi + o_{P}(\boldsymbol{1}_{p-1}),\quad \bxi \mid \M(\V) \sim N_{p-1}(\boldsymbol{0}, \s^2\bSigma^{-1/2}\widehat\bOmega\bSigma^{-1/2}).
    \end{equation}
Here, $\widehat\bOmega=\M(\V)^\top \M(\V)/n$ is the sample covariance of $\M(\V)$. Note that $o_{P}(\boldsymbol{1}_{p-1})$ is a random vector satisfying $\|o_{P}(\boldsymbol{1}_{p-1})\|_{\infty}\to 0$ in probability as $n,p\to\infty$. 
Thus, the asymptotic covariance of $\widehat\balpha$ is $\s^2\bSigma^{-1/2}\widehat\bOmega\bSigma^{-1/2}/n$.
\end{proposition}
Define the residual $\bDelta=\sqrt{n}(\widehat\balpha-\balpha) - \bxi\in\R^{p-1}$, we also derive that under the settings of Proposition~\ref{prop:Proposed}, if the sample size satisfies $n \ge s_0\log (p-1)$, then the maximum size of the residual $\Delta_j$ over $j\in\{1,\ldots,p-1\}$ is bounded by $\|\bDelta\|_{\infty} = O_P\left(\sqrt{\frac{s_0}{n}}\log (p-1)\right)$. Proofs and technical details are provided in the Appendix Section~3.

For any positive constant $A$, let $\mathcal{G}_n=\mathcal{G}_n(A)$ be the event that
\[
    \mathcal{G}_n(A) = \left\{\M(\V)\in\R^{n\times(p-1)}:\,  |(\bSigma^{-1/2}\widehat\bOmega\bSigma^{-1/2})\bSigma - \I_{p-1}|_{\infty}\le A\sqrt{\frac{\log (p-1)}{n}}\right\}.
\]
We provide the lower bound of the asymptotic variance of $\widehat\balpha$ on the event $\mathcal{G}_n(A)$ in Lemma~\ref{lem:lowerbound}.
\begin{lemma}\label{lem:lowerbound}
    Assume that $\sigma_{\min}(\bSigma)\ge C_{\min}>0$ and $\sigma_{\max}(\bSigma)\le C_{\max}<\infty$, denote $A(n,p) = A\sqrt{(\log (p-1))/n}$. Then, we have
    $
        P(\M(\V)\in \mathcal{G}_n(A))\ge 1 - 2(p-1)^{-2},
    $
    and for $j=1,\ldots,p-1$, on the event $\mathcal{G}_n(A)$,
    $  
        [\bSigma^{-1/2}\widehat\bOmega\bSigma^{-1/2}]_{j,j}\ge \frac{(1-A(n,p))^2}{\widehat\bSigma_{j,j}}.
    $
\end{lemma}

\begin{theorem}\label{thm:Proposed}
    Under the assumptions of Proposition~\ref{prop:Proposed} and Lemma~\ref{lem:lowerbound}, if $s_0\ll n/(\log (p-1))^2$, then $\widehat\balpha$ is normally distributed. More precisely, let $\widehat\sigma$ be an estimator of $\sigma$ satisfying
    \begin{eqnarray}\label{eq:sigmahat}
        \lim_{n\to \infty}\sup_{\balpha\in\R^{p-1},\|\balpha\|_0\le s_0} P\left(\left| \frac{\widehat\sigma}{\sigma}-1\right|\ge \epsilon\right)=0, \quad \text{for any $\eps>0$}.
    \end{eqnarray}
    If $s_0\ll n/(\log (p-1))^2$, then, for any $x\in \R$, we have the following almost surely
    \begin{eqnarray}\label{eq:a.s.normal}
        \lim_{n\to\infty} \sup_{\balpha\in\R^{p-1};\|\balpha\|_0\le s_0}\left| P\left(\frac{\sqrt{n}(\widehat\alpha_j-\alpha_j)}{\widehat\sigma[\bSigma^{-1/2}\widehat\bOmega\bSigma^{-1/2}]^{1/2}_{j,j}}\le x\right)-\Phi(x)\right| = 0,
    \end{eqnarray}
    where $\Phi(x)$ is the cumulative distribution function of the standard normal distribution.
\end{theorem}

Existing studies suggest various ways of obtaining a consistent estimator of the noise level $\sigma$ \citep{fan2001variable,fan2008sure,sun2012scaled}. For demonstration, we use the scaled Lasso \citep{sun2012scaled} given by
\begin{eqnarray}\label{eq:scaleDeLasso}
    \{\widehat\balpha^*,\widehat\sigma\} = \arg\min_{\balpha\in \R^{p-1},\sigma>0} \left\{\frac{1}{2n\sigma}\|\y-\M(\V)\balpha\|^2_2 + \frac{\sigma}{2} + \widetilde\lambda\|\balpha\|_1\right\}.
\end{eqnarray}
With $\widetilde\lambda=10\sqrt{(2\log (p-1))/n}$, the estimator $\widehat\sigma$ satisfies Eq~\eqref{eq:sigmahat} by the analyses of \citet{sun2012scaled} and \citet{javanmard2014confidence}.

Based on the distribution characterization of $\widehat\balpha$ given by Eq~\eqref{eq:a.s.normal}, it is quite straightforward to construct asymptotically valid confidence intervals for each parameter $\alpha_j$. Namely, for $j=1,\ldots,p-1$ and significant level $a\in (0,1)$, the confidence interval of $\alpha_j$ is
\begin{eqnarray}
    J_j(a) &=& [\widehat\alpha_j - \delta(a,n), \widehat\alpha_j + \delta(a,n)],\label{eq:ci}\\
    \delta(a,n) &=& \Phi^{-1}\left(1-\frac{a}{2}\right)\frac{\widehat\sigma}{\sqrt{n}}\left(\bSigma^{-1/2}\widehat\bOmega\bSigma^{-1/2}\right)^{1/2}_{j,j}.\notag
\end{eqnarray}

\begin{corollary}\label{cor:ci}
    The confidence interval $J_j(a)$ of $\alpha_j$, defined by Eq~\eqref{eq:ci}, is asymptotically valid, that is,
    $
        \lim_{n\to\infty} P(\alpha_j\in J_j(a)) = 1-a.
    $
\end{corollary}
Finally, for hypothesis testing, we consider the null hypothesis $H_{0,j}:\alpha_j=0$ versus the alternative $H_{1,j}:\alpha_j\ne 0$. We can construct a two sided $p$-value for testing $H_{0,j}$ as follows:
$
    p_j = 2-2\Phi\left(\sqrt{n}| \widehat\alpha_j|/(\widehat\sigma[\bSigma^{-1/2}\widehat\bOmega\bSigma^{-1/2}]^{1/2}_{j,j})\right).
$

Our theoretical results are established based on known nuisance parameters $\s^2_u,\bmu_{\widetilde z}$, and $\bSigma_{\widetilde z}$, which yield deterministic covariates $\M(\V)$ and its covariance matrix $\bSigma$ given the observed data $\V$. With estimated $\widehat\s^2_u,\widehat\bmu_{\widetilde z},\widehat\bSigma_{\widetilde z}$, we can obtain $\widehat{\M}(\V)$ with the $i$th row given by $\widehat\bmu(\V_i) = \widehat\bmu_{\widetilde z} + \widehat\bSigma_{\widetilde z} 
(\widehat\bSigma_{\widetilde z}+2\widehat\s^2_u\I_{p-1})^{-1}(\widetilde{\V}_{i,-p} - \widehat\bmu_{\widetilde z})$, and $\widehat\bSigma = \widehat\bSigma_{\widetilde z} (\widehat\bSigma_{\widetilde z}+2\widehat\s^2_u\I_{p-1})^{-1}\widehat\bSigma_{\widetilde z}$ based on Eqs~\eqref{eq:rcnormal}-\eqref{eq:Sigma}, respectively. Consequently, the proposed debiased-Lasso type estimator can be written as
$
    \widehat\balpha =  \widetilde\balpha +(\widehat\bSigma)^{-1}\widehat{\M}(\V)^\t (\y -\widehat{\M}(\V)\widetilde\balpha)/n,
$
where $\widetilde\balpha = \widetilde\balpha(\y,\widehat{\M}(\V);\lambda)$ is the Lasso estimator of $\y$ regressing on $\widehat{\M}(\V)$ with the regularization parameter $\lambda$. Deriving the asymptotic normality of $\widehat\balpha$ is challenging because $\widehat{\M}(\V)$ and $\widehat\bSigma$ nonlinearly depend on the estimated nuisance parameters, leading to the absence of an explicit relationship between the estimated $\widehat{\M}(\V),\widehat\bSigma$ and the true $\M(\V),\bSigma$.  In contrast to the traditional debiased-Lasso estimator, which deals with only the unknown covariance matrix of the deterministic covariates \citep{javanmard2018debiasing}, our setting involves multiple nonlinear dependencies due to measurement errors. However, when $|\widehat{\M}(\V) - \M(\V)|_{\infty}\to 0$ and $|\widehat\bSigma - \bSigma|_{\infty}\to 0$, the asymptotic normality of $\widehat\balpha$ still holds. Further details and rigorous proof are discussed in Appendix Section~3.7
, with the latter warranted for future work.

Much of the literature about debiased-Lasso type methods or measurement error problems, such as \citet{xu2011goodness} and \citet{van2019asymptotic}, similarly considered nuisance parameters to be known. In the context of microbiome studies, nuisance parameters $\s^2_u,\bmu_{\widetilde z},\bSigma_{\widetilde z}$ are fully determined by $\s^2_u,\bmu_x,\bSigma_x$, with their estimators provided in Section~\ref{sec:implement}. We can obtain consistent estimators for $\s^2_u$ and $\bmu_x$ using Eq~\eqref{eq:sigma_u} and Eq~\eqref{eq:mu_x}, respectively. For $\bSigma_x$, the node-wise estimator $\widehat\bSigma_{x,nw}$ defined in Eq~\eqref{eq:nodewise} is a Lasso-type estimator based on the Gaussian graphical model \citep[Chapter 5.2]{lauritzen1996graphical}, which facilitates the derivation of desired asymptotic properties of the proposed estimator $\widehat\balpha$ based on the theory of the debiased Lasso, such as \citet[Theorem 1.3]{javanmard2018debiasing}. The shrinkage estimator $\widehat\bSigma_{x,sh}$ defined in Eq~\eqref{eq:shrinkage} is shrunk towards the diagonal matrix, leading to a bias of non-diagonal elements, not desirable regarding deriving asymptotic properties of the proposed estimator $\widehat\balpha$. Although the comprehensive analysis of nuisance parameter estimation on theoretical results falls outside this paper's scope, we recognize it as a notable challenge in high-dimensional data analysis, meriting further exploration in subsequent studies.

\section{Simulation Studies}\label{sec:simulation}
To evaluate the performance of the proposed high-dimensional calibration method, we conduct simulation studies under different settings. We compare our method with Lasso, debiased Lasso, and CoCoLasso with data generated by models \eqref{eq:logcontrastV} and \eqref{eq:mem_composition}. 
To simulate the true compositional data matrix $\Z=(z_{ij})$, we set $(n,p)=(100,100)$ and $(200,300)$, and generate $\X$ based on $\log(\X_i)\sim N_p(\bmu_x,\bSigma_x)$ with $\bmu_x=(\mu_1,\ldots,\mu_p)^\t$ and $\bSigma_x=(\sigma_{ij})\in\R^{p\times p}$. Here, we set $\mu_j=\log(p/2)$ for $j=1,\ldots,5$ and $\mu_j=0$ otherwise to allow component proportions to differ by orders of magnitude, as is the case for the gut microbiome data. To describe correlations among the components, we take $\sigma_{ij}=\rho^{|i-j|}$ with $\rho=0.2$. Then, we transform $\X$ to $\Z$ by $z_{ij} = x_{ij}/(\sumjp x_{ij})$. We generate $\W=(w_{ij})$ based on Eq~\eqref{eq:mem} with $\s^2_u=1$ and then transform $\W$ to the compositional data matrix with measurement error $\V=(v_{ij})$ by $v_{ij} = w_{ij}/(\sumjp w_{ij})$. We also generate three additional replicates of $\W$ to identify the variance of the measurement error $\s^2_u$ based on Eq~\eqref{eq:sigma_u}. The response $y_i$ is generated based on Eq~\eqref{eq:logcontrastZ} with the true coefficient vector $\balpha = (1,-0.8,1.5,0.6,-0.9,1.2,0.4,0,\ldots,0)^\t\in\R^{p-1}$ and error term $\eps_i\simiid N(0,0.5^2)$ for $i=1,\ldots,n$.

We estimate the noise level $\sigma$ by the scaled Lasso based on Eq~\eqref{eq:scaleDeLasso}.
Additionally, we estimate $\s^2_u$ based on Eq~\eqref{eq:sigma_u} and estimate $\bmu_x$ based on Eq~\eqref{eq:mu_x}, and we estimate $\bSigma_x$ by shrinkage estimator $\widehat\bSigma_{x,sh}$ based on Eq~\eqref{eq:shrinkage} since the sample size ($n$) is small in the context of high dimensional data. For the Lasso, the debiased Lasso, and the proposed method, we use five-fold cross-validation to select the tuning parameter $\lambda$. Given true nuisance parameters $\s^2_u,\bmu_x,\bSigma_x$, as well as estimated $\widehat\s^2_u,\widehat\bmu_x,\widehat\bSigma_{x,sh}$, we report the empirical average bias, root mean square error, standard error, and coverage rate of the nominal 95\% confidence interval for the debiased Lasso method and the proposed method based on $N=500$ Monte Carlo replicates. Since neither of Lasso nor CoCoLasso has a variance estimation theory, we only report the empirical average bias based on $N=500$ Monte Carlo replicates. We present the statistical inference results for $\alpha_1,\ldots,\alpha_{10}$ in Tables \ref{tab:n200rho0.2}-\ref{tab:n100rho0.2}. 

\begin{table}
\begin{center}
\caption{Results for $n=200,p=300$ with true and estimated $\s^2_u,\bmu_x,\bSigma_x$. Bias, average bias; RMSE, root mean square error; SE, standard error; CR, coverage rate of the 95\% confidence interval; Coco, CoCoLasso method; DeLasso, debiased lasso method; Proposed, proposed high-dimensional calibration method.}\label{tab:n200rho0.2}
\resizebox{\columnwidth}{!}{
\begin{tabular}{ccccccccccc}
\hline
\multicolumn{11}{c}{With true $\s^2_u,\bmu_x,\bSigma_x$:}\\
\hline
& \multicolumn{4}{c}{\bfseries \normalsize Bias} & \multicolumn{2}{c}{\bfseries \normalsize RMSE} & \multicolumn{2}{c}{\bfseries \normalsize SE} & \multicolumn{2}{c}{\bfseries \normalsize CR} \\
\cmidrule(r){2-5} \cmidrule(r){6-7} \cmidrule(r){8-9} \cmidrule(r){10-11}
True $\alpha_j$ & Lasso & Coco & DeLasso & Proposed & DeLasso & Proposed & DeLasso & Proposed & DeLasso & Proposed\\
\hline
  1.00 & -0.81 & -0.78 &-0.55 & -0.04 & 0.57 & 0.31 & 0.13 & 0.30 & 0.02 & 0.94 \\ 
 -0.80 &  0.78 &  0.76 & 0.55 &  0.05 & 0.57 & 0.32 & 0.13 & 0.31 & 0.04 & 0.95 \\ 
  1.50 & -1.05 & -0.82 &-0.81 & -0.07 & 0.82 & 0.33 & 0.13 & 0.31 & 0.00 & 0.92 \\ 
  0.60 & -0.50 & -0.50 &-0.29 & -0.03 & 0.33 & 0.32 & 0.13 & 0.31 & 0.37 & 0.93 \\ 
 -0.90 &  0.86 &  0.84 & 0.58 &  0.07 & 0.60 & 0.32 & 0.13 & 0.31 & 0.01 & 0.94 \\ 
  1.20 & -0.90 & -0.81 &-0.65 & -0.05 & 0.67 & 0.33 & 0.13 & 0.31 & 0.00 & 0.92 \\ 
  0.40 & -0.34 & -0.34 &-0.17 & -0.03 & 0.22 & 0.32 & 0.13 & 0.31 & 0.68 & 0.94 \\ 
  0.00 &  0.00 &  0.00 & 0.03 &  0.01 & 0.15 & 0.32 & 0.13 & 0.31 & 0.91 & 0.95 \\ 
  0.00 &  0.00 &  0.00 & 0.00 &  0.00 & 0.14 & 0.31 & 0.13 & 0.31 & 0.92 & 0.95 \\ 
  0.00 &  0.00 &  0.00 & 0.00 &  0.00 & 0.13 & 0.29 & 0.13 & 0.31 & 0.92 & 0.95 \\
\hline
\multicolumn{11}{c}{with estimated $\widehat\s^2_u,\widehat\bmu_x,\widehat\bSigma_{x,sh}$:}\\
\hline
& \multicolumn{4}{c}{\bfseries \normalsize Bias} & \multicolumn{2}{c}{\bfseries \normalsize RMSE} & \multicolumn{2}{c}{\bfseries \normalsize SE} & \multicolumn{2}{c}{\bfseries \normalsize CR} \\
\cmidrule(r){2-5} \cmidrule(r){6-7} \cmidrule(r){8-9} \cmidrule(r){10-11}
True $\alpha_j$ & Lasso & Coco & DeLasso & Proposed & DeLasso & Proposed & DeLasso & Proposed & DeLasso & Proposed\\
\hline
   1.00  & -0.81 & -0.78 & -0.58 & -0.18 & 0.60 & 0.35 & 0.12 & 0.27 & 0.01 & 0.86 \\ 
  -0.80  &  0.77 &  0.76 &  0.60 &  0.41 & 0.62 & 0.50 & 0.12 & 0.27 & 0.01 & 0.62 \\ 
   1.50  & -1.04 & -0.82 & -0.80 & -0.15 & 0.82 & 0.34 & 0.12 & 0.27 & 0.00 & 0.88 \\ 
   0.60  & -0.50 & -0.50 & -0.29 &  0.00 & 0.33 & 0.30 & 0.12 & 0.27 & 0.37 & 0.94 \\ 
  -0.90  &  0.85 &  0.84 &  0.63 &  0.38 & 0.65 & 0.47 & 0.12 & 0.27 & 0.00 & 0.68 \\ 
   1.20  & -0.90 & -0.81 & -0.67 & -0.17 & 0.69 & 0.34 & 0.12 & 0.27 & 0.00 & 0.86 \\ 
   0.40  & -0.33 & -0.34 & -0.15 &  0.08 & 0.21 & 0.29 & 0.12 & 0.27 & 0.68 & 0.92 \\ 
   0.00  &  0.00 &  0.00 &  0.04 &  0.08 & 0.15 & 0.29 & 0.12 & 0.27 & 0.89 & 0.93 \\ 
   0.00  &  0.00 &  0.00 &  0.00 &  0.00 & 0.13 & 0.26 & 0.12 & 0.27 & 0.95 & 0.96 \\ 
   0.00  &  0.00 &  0.00 &  0.00 &  0.00 & 0.13 & 0.25 & 0.12 & 0.27 & 0.92 & 0.96 \\
\hline
\end{tabular}
}
\end{center}
\end{table}

\begin{table}
\begin{center}
\caption{Results for $n=p=100$ with true and estimated $\s^2_u,\bmu_x,\bSigma_x$. Bias, average bias; RMSE, root mean square error; SE, standard error; CR, coverage rate of the 95\% confidence interval; Coco, CoCoLasso method; DeLasso, debiased lasso method; Proposed, proposed high-dimensional calibration method.\label{tab:n100rho0.2}}
\resizebox{\columnwidth}{!}{
\begin{tabular}{ccccccccccc}
\hline
\multicolumn{11}{c}{With true $\s^2_u,\bmu_x,\bSigma_x$:}\\
\hline
& \multicolumn{4}{c}{\bfseries \normalsize Bias} & \multicolumn{2}{c}{\bfseries \normalsize RMSE} & \multicolumn{2}{c}{\bfseries \normalsize SE} & \multicolumn{2}{c}{\bfseries \normalsize CR} \\
\cmidrule(r){2-5} \cmidrule(r){6-7} \cmidrule(r){8-9} \cmidrule(r){10-11}
True $\alpha_j$ & Lasso & Coco & DeLasso & Proposed & DeLasso & Proposed & DeLasso & Proposed & DeLasso & Proposed\\
\hline
   1.00 & -0.83 & -0.81 & -0.58 & -0.10 & 0.62 & 0.45 & 0.17 & 0.41 & 0.13 & 0.92 \\ 
  -0.80 &  0.78 &  0.77 &  0.57 &  0.11 & 0.60 & 0.43 & 0.17 & 0.43 & 0.13 & 0.93 \\ 
   1.50 & -1.09 & -0.92 & -0.83 & -0.12 & 0.86 & 0.51 & 0.17 & 0.43 & 0.02 & 0.89 \\ 
   0.60 & -0.50 & -0.50 & -0.30 & -0.04 & 0.36 & 0.45 & 0.17 & 0.43 & 0.53 & 0.94 \\ 
  -0.90 &  0.86 &  0.86 &  0.60 &  0.11 & 0.63 & 0.44 & 0.18 & 0.43 & 0.12 & 0.95 \\ 
   1.20 & -0.95 & -0.88 & -0.69 & -0.13 & 0.73 & 0.48 & 0.17 & 0.43 & 0.06 & 0.92 \\ 
   0.40 & -0.33 & -0.33 & -0.17 & -0.02 & 0.27 & 0.44 & 0.17 & 0.43 & 0.77 & 0.95 \\ 
   0.00 &  0.01 &  0.00 &  0.02 &  0.01 & 0.18 & 0.39 & 0.17 & 0.43 & 0.96 & 0.98 \\ 
   0.00 &  0.00 &  0.00 & -0.01 & -0.02 & 0.19 & 0.41 & 0.17 & 0.43 & 0.92 & 0.95 \\ 
   0.00 &  0.01 &  0.01 &  0.01 &  0.04 & 0.19 & 0.41 & 0.17 & 0.43 & 0.92 & 0.96 \\
\hline
\multicolumn{11}{c}{with estimated $\widehat\s^2_u,\widehat\bmu_x,\widehat\bSigma_{x,sh}$:}\\
\hline
& \multicolumn{4}{c}{\bfseries \normalsize Bias} & \multicolumn{2}{c}{\bfseries \normalsize RMSE} & \multicolumn{2}{c}{\bfseries \normalsize SE} & \multicolumn{2}{c}{\bfseries \normalsize CR} \\
\cmidrule(r){2-5} \cmidrule(r){6-7} \cmidrule(r){8-9} \cmidrule(r){10-11}
True $\alpha_j$ & Lasso & Coco & DeLasso & Proposed & DeLasso & Proposed & DeLasso & Proposed & DeLasso & Proposed\\
\hline
   1.00  & -0.81 & -0.81  & -0.58 & -0.19 & 0.62 & 0.45 & 0.17 & 0.37 & 0.15 & 0.89 \\ 
  -0.80  &  0.78 &  0.77  &  0.62 &  0.45 & 0.65 & 0.57 & 0.17 & 0.37 & 0.07 & 0.77 \\ 
   1.50  & -1.07 & -0.92  & -0.81 & -0.19 & 0.84 & 0.47 & 0.17 & 0.37 & 0.02 & 0.84 \\ 
   0.60  & -0.49 & -0.50  & -0.28 &  0.00 & 0.35 & 0.38 & 0.17 & 0.37 & 0.57 & 0.95 \\ 
  -0.90  &  0.86 &  0.86  &  0.65 &  0.43 & 0.68 & 0.57 & 0.17 & 0.37 & 0.09 & 0.75 \\ 
   1.20  & -0.93 & -0.88  & -0.68 & -0.22 & 0.71 & 0.47 & 0.17 & 0.37 & 0.05 & 0.86 \\ 
   0.40  & -0.32 & -0.33  & -0.14 &  0.09 & 0.24 & 0.40 & 0.17 & 0.37 & 0.80 & 0.93 \\ 
   0.00  &  0.01 &  0.00  &  0.05 &  0.09 & 0.19 & 0.35 & 0.17 & 0.37 & 0.92 & 0.96 \\ 
   0.00  &  0.00 &  0.00  &  0.00 &  0.00 & 0.18 & 0.35 & 0.17 & 0.37 & 0.92 & 0.95 \\ 
   0.00  &  0.00 &  0.01  &  0.00 &  0.01 & 0.18 & 0.35 & 0.17 & 0.37 & 0.93 & 0.96 \\ 
\hline
\end{tabular}
}
\end{center}
\end{table}

As seen from Tables \ref{tab:n200rho0.2}-\ref{tab:n100rho0.2}, the Lasso estimator, due to the neglect of measurement errors, has inferior performance in all settings. CoCoLasso, taking measurement errors into account, also performs poorly in terms of point estimation in our challenging situations with small $n$, large $p$, and large measurement errors. This is because CoCoLasso, as a modification of the Lasso, focuses on improving prediction accuracy and model interpretability at the expense of a bias. Therefore, we specifically compare our proposed method with the debiased Lasso. 

We first assess the estimation results for non-zero $\alpha_j$. For the point estimator, we observe that the proposed estimator exhibits a significantly smaller bias than the debiased Lasso estimator. When using the estimated $\widehat\s^2_u,\widehat\bmu_x,\widehat\bSigma_{x,sh}$, our proposed estimator still has slight biases for some $\alpha_j$ since $\widehat\bSigma_{x,sh}$ is biased. However, with true $\s^2_u,\bmu_x,\bSigma_x$, the bias of the proposed estimator approaches zero, and the root mean square error is smaller than that of the debiased Lasso estimator. For the variance estimator, our method usually has a larger standard error than the debiased Lasso due to the variance-bias trade-off. The increase in standard error caused by correcting measurement errors has been reported in the existing literature; see \citet{achic2018categorizing} and \citet[Chapter 9.4]{carroll2006measurement}. Regarding the coverage rate, the debiased Lasso has poor coverage rates across all cases (all less than 80\% and close to zero for large $\alpha_j$). However, our method maintains a nominal coverage rate with true $\s^2_u,\bmu_x,\bSigma_x$ in most cases. In the most challenging cases, where we need to estimate $\s^2_u,\bmu_x$, and $\bSigma_{x,sh}$, our method has acceptable confidence intervals (mostly more than 85\%), whereas the confidence intervals of debiased Lasso estimator are lower than 40\% coverage. This suggests that our method is relatively robust despite the minor under-coverage issue. In microbiome studies, the small sample size $n$ could lead to inaccurate estimation of the high-dimensional covariance matrix $\bSigma_x$, which affects the performance on debiased Lasso-type methods. However, the deficiency in the sample size does not affect the core of our methodology and is expected to be addressed with the development of technology. 

For the simpler case ($\alpha_j=0$), all methods we compared have close-to-zero bias, and the coverage rates of confidence intervals based on the debiased Lasso and our method are similar, indicating that the Lasso-type methods can also identify zero entries of $\balpha$ and conduct variable selection even if ignoring the measurement errors. 

We also conducted additional simulations for $n=500,p=600$ under a more challenging setting with $\rho=0.5,\s^2_u=0.5$. Results for both the true and estimated nuisance parameters $\s^2_u,\bmu_x,\bSigma_x$ cases are consistent with the above simulations. Detailed results are shown in Table~A2 
in Appendix Section~4.

\section{Data Analysis}\label{sec:data}
In this section, we apply the proposed high-dimensional calibration method to a longitudinal microbiome study reported by \citet{flores2014temporal}, which has also been studied for investigating the association between microbiome and traits \citep{shi2022high}. Our focus is on the association between BMI and gut microbiome composition at the genus level for healthy adults. We consider $p=40$ bacteria genera appear in more than 60\% of the original 352 samples of $n=41$ subjects with BMI recorded and who did not have antibiotics. Within each of the 41 subjects, we exclude samples that deviate significantly from the others and retain the four closest measurements as repeated observations within a three-month period. This selection is based on the dissimilarity indices for community ecologists, computed using the \texttt{vegdist} function from the \texttt{R} package \texttt{vegan} \citep{shi2022high}. 

To implement the proposed high-dimensional calibration estimator, we first perform the classical zero-imputation method in the literature \citep{lubbe2021comparison,shi2022high} with zero counts replaced to a small constant (e.g., 0.1). The Kolmogorov-Smirnov test does not show significant differences for the 40 components before and after the zero-replacement. We treat four observations for each subject within three months as repeated measurements to identify the measurement error. Then we estimate $\sigma^2_u$ based on $n=41$ subjects with four repeated recalls and 15 bacteria genera, which have 90\% nonzero proportion in the original observed count data. The estimated $\widehat\sigma^2_u=1.16$. Since our simulations have demonstrated that the Lasso estimator and the CoCoLasso estimator are inferior in all respects, we only compare our method with the debiased Lasso. Regarding \textit{Blautia} as $p$-th genus, we run the debiased Lasso and the proposed method with 10-fold cross-validation choosing the tuning parameter $\lambda$. We consider the $j$th genus to be associated with BMI if the $p$-value for testing $\alpha_j=0$ is less than the significant level of 0.05. We report the point estimator, standard error, $p$-value, and the nominal 95\% confidence interval results for the significant genera selected by either the debiased Lasso estimator or the proposed estimator in Table \ref{tab:dataanalysis1}. Results for other insignificant genera are presented in Table~A3; 
see Appendix Section~5. 

Table~\ref{tab:dataanalysis1} shows the point estimator of $\alpha_j$ from the proposed method is greater than the debiased Lasso, indicating that the debiased Lasso underestimates the effects of genera on BMI. The standard error of the proposed method is greater than the debiased Lasso, leading to a wider confidence interval, which is more trustworthy based on our simulation results (Tables~\ref{tab:n200rho0.2}-\ref{tab:n100rho0.2}). The genera selected by both methods are \textit{Dialister}, \textit{Finegoldia}, \textit{Lachnospira}, \textit{Lactobacillus}, \textit{Parabacteroides} and \textit{Porphyromonas}. Though both the debiased Lasso and the proposed method identified the same six genera, it is due to the coincidence that the debiased Lasso yields biased point estimation far from zero and underestimated variance, resulting in ``significant" $p$-values. \textit{Varibaculum} is only selected by our proposed method, which has been reported to be related to hepatic fat \citep{stanislawski2018gut}. Subsequent analyses between response and \textit{Varibaculum} also indicate that subjects with a higher level of \textit{Varibaculum} tend to have a higher BMI; see Appendix Section~5 
for details. 

\begin{table}
\begin{center}
\caption{Data analysis results. DeLasso, debiased lasso method; Proposed, proposed high-dimensional calibration method; Estimate, point estimator; SE, standard error; $p$-value, $p$-value of hypothesis testing $H_0:\alpha_j=0$; 95\% CI, nominal 95\% confidence interval of $\alpha_j$.}\label{tab:dataanalysis1}
\resizebox{\columnwidth}{!}{
\begin{tabular}{lcccccccc}
\hline
& \multicolumn{4}{c}{\bfseries \normalsize DeLasso} & \multicolumn{4}{c}{\bfseries \normalsize Proposed} \\
\cmidrule(r){2-5} \cmidrule(r){6-9}
Genus & Estimate & SE & $p$-value & 95\% CI & Estimate & SE & $p$-value & 95\% CI\\
\hline
  Dialister       &  0.21 & 0.08 & 0.01 & ( 0.05,  0.38) &  0.24  & 0.11 & 0.02 &( 0.03, 0.46)\\ 
  Finegoldia      &  0.24 & 0.08 & 0.00 & ( 0.08,  0.41) &  0.51  & 0.16 & 0.00 &( 0.21, 0.82)\\ 
  Lachnospira     & -0.23 & 0.12 & 0.04 & (-0.46, -0.01) & -0.55  & 0.21 & 0.01 &(-0.97,-0.13) \\ 
  Lactobacillus   & -0.27 & 0.09 & 0.00 & (-0.46, -0.09) & -0.43  & 0.13 & 0.00 &(-0.68,-0.17) \\ 
  Parabacteroides &  0.27 & 0.12 & 0.02 & ( 0.04,  0.50) &  0.40  & 0.19 & 0.03 &( 0.04, 0.77)\\ 
  Porphyromonas   & -0.27 & 0.08 & 0.00 & (-0.44, -0.11) & -0.35  & 0.16 & 0.02 &(-0.66,-0.05) \\ 
  Varibaculum     &  0.16 & 0.09 & 0.08 & (-0.02,  0.34) &  0.42  & 0.18 & 0.02 &( 0.07, 0.76) \\
\hline
\end{tabular}
}
\end{center}
\end{table}

\section{Discussion}\label{sec:conclusion}
This paper introduces an innovative approach for conducting inference on contaminated high-dimensional compositional data. While our examples primarily focus on compositional data, the methodology is equally applicable to general high-dimensional regression problems as discussed in the literature \citep{loh2012high,datta2017cocolasso}. Given that compositional data often originate from count data, there is a potential for this method to be adapted for correcting contaminated count data that contain measurement errors.

We noted a slight bias that arises from estimating nuisance parameters with a limited number of samples, a point we touch upon in Section~\ref{sec:simulation}. While the estimation of these parameters is not the central focus of our paper, it is worth noting that there is an extensive corpus of research dedicated to the estimation of covariance matrices in high-dimensional contexts \citep{bickel2008regularized,cai2012optimal,van2014asymptotically}. Insights from these studies could potentially enhance the efficacy of our proposed method in future applications. 

This work marks an initial effort to explore statistical inference within the realm of high-dimensional contaminated datasets, extending beyond the scope of microbiome data. We have not delved deeply into the issue of zero inflation, where an excessive number of zeros in the data may not conform to conventional distribution models. The phenomenon of zero inflation, particularly when some zeros might result from measurement errors, presents an intriguing area for further investigation.  In addition, raw RNA gene sequences are often clustered into operational taxonomic units at the species level in human microbiome studies \citep{chen2013structure,wang2017structured}. Integrating information from phylogenetic trees could enhance variable selection, offering a promising avenue for expanding the application of the proposed method in future research endeavors.


\begin{center}
{\large\bf SUPPLEMENTARY MATERIAL}
\end{center}

The supplemental materials provide a summary of notations, mathematical details of high-dimensional regression with errors-in-variables, proofs of all technical details, additional simulations, and additional results on data analysis.




\bibliographystyle{apalike}
\bibliography{ref}

\end{document}